\documentclass{an}
\usepackage{graphicx}
\usepackage{times}
\newcommand{\kms}{km~s$^{-1}$}

\begin{document}
\headnote{AN 322 (2001) 5/6, 411--418}

\title{Search for progenitors of supernovae type Ia with SPY\thanks{Based on 
data obtained at the Paranal Observatory of the European Southern Observatory 
for programs 165.H-0588, 167.D-0407, and 266.D-5658}
\fnmsep
\thanks{Based on observations collected at the German-Spanish Astronomical 
Center, Calar Alto, operated by the Max-Planck-Institut f\"ur Astronomie 
Heidelberg jointly with the Spanish National Commission for Astronomy}}

\author{R.~Napiwotzki\inst{1}
\and N.~Christlieb\inst{2}
\and H.~Drechsel\inst{1}
\and H.-J.~Hagen\inst{2}
\and U.~Heber\inst{1}
\and D.~Homeier\inst{3}\fnmsep\inst{4}
\and C.~Karl\inst{1}
\and D.~Koester\inst{3}
\and B.~Leibundgut\inst{5}
\and T.R.~Marsh\inst{6}
\and S.~Moehler\inst{1}
\and G.~Nelemans\inst{7}\fnmsep\inst{8}
\and E.-M.~Pauli\inst{1}
\and D.~Reimers\inst{2}
\and A.~Renzini\inst{5}
\and L.~Yungelson\inst{9}}

\institute{Dr.~Remeis-Sternwarte, Astronom.\ Institut, Universit\"at 
	Erlangen-N\"urnberg, Sternwartstr.~7, 96049 Bamberg, Germany
\and 
Hamburger Sternwarte, Universit\"at Hamburg, Gojenbergsweg 112, 21029 Hamburg,
	Germany
\and
Institut f\"ur Theoretische Physik und Astrophysik, Universit\"at Kiel, 
	24098 Kiel, Germany
\and
Department of Physics \& Astronomy,
	University of Georgia, Athens, GA\,30602-2451, USA
\and
European Southern Observatory, Karl-Schwarzschild-Str.~2, 85748 Garching, 
	Germany
\and
University of Southampton, Department of Physics \& Astronomy, Highfield,
	Southampton S017 1BJ, UK
\and 
Astronomical Institute ``Anton Pannekoek'', Kruislaan 400, 1098\,SJ Amsterdam,
	The Netherlands
\and 
Institute of Astronomy, Madingley Road, Cambridge CB3~0HA, UK
\and
Institute of Astronomy of the Russian Academy of Sciences, 48 Pyatnitskaya 
	Str., 109017 Moscow, Russia
}

\date{Received {\it date};
accepted {\it date}}

\abstract{We have started a large survey for double degenerate (DD)
binaries as potential progenitors of type Ia supernovae with the UVES
spectrograph at the ESO VLT (ESO {\bf S}N\,Ia {\bf P}rogenitor
surve{\bf Y} -- SPY). About 400 white dwarfs were checked for radial
velocity variations during the first 15 months of this project, twice
the number of white dwarfs investigated during the last 20 years. We
give an overview of the SPY project and present first results 
Fifty four new DDs have been discovered, seven of them double lined
(only 18 and 6 objects of these groups were known before, respectively). 
The final sample is
expected to contain 150 to 200 DDs. Eight new pre-cataclysmic binaries
were also detected. SPY is the first DD survey which encompasses also
non-DA white dwarfs.  SPY produces an immense, unique sample of very
high resolution white dwarf spectra, which provides a lot of spin-off
opportunities.  We describe our projects to exploit the SPY sample for
the determination of basic parameters, kinematics, and rotational
velocities of white dwarfs. A catalogue with a first subset of our white
dwarf data has already been published by Koester et al.\
(\cite{KNC01}).}

\maketitle

\section{Introduction}

Supernovae of type Ia (SN\,Ia) play an outstanding role for our
understanding of galactic evolution and the determination of the
extragalactic distance scale.  However, the nature of their
progenitors is still unknown (e.g.\ Livio \cite{Liv00}).  There is general
consensus that the event is due to the thermonuclear explosion of a
white dwarf when a critical mass (very likely the Chandrasekhar limit,
$1.4M_\odot$) is reached, but the nature of the progenitor system
remains unclear. It must be a binary, with matter being transfered to
the white dwarf from a companion until the critical mass is
reached. However, two options exist for the nature of the companion:
either another white dwarf in the so-called double degenerate ({\bf
DD}) scenario (Iben \& Tutukov \cite{IT84}), or a red giant/subgiant
in the so-called single degenerate ({\bf SD}) scenario (Whelan \& Iben
\cite{WI73}), with the system possibly appearing as a symbiotic binary
(Munari \& Renzini \cite{MR92}) or a supersoft X-ray source (van den
Heuvel et al.\ \cite{vHBN92}).

The solution of the SD vs.\ DD dilemma is of great importance for
assessing the role of SNe~Ia in a variety of astrophysical situations,
and -- perhaps even more importantly -- their effectiveness as
accurate cosmological probes (cf.\ Leibundgut \cite{Lei00}).  
As is widely known, current studies of
high-$z$ SNe~Ia favour an {\it accelerating}, $\Lambda$--dominated
universe (Riess et al.\ \cite{RFC98}; Leibundgut \cite{Lei01}).  
Yet, the result
depends on the assumption that the light-curves of high-$z$ SNe~Ia are
intrinsically identical to those of the local SNe~Ia. If instead
high-$z$ SNe~Ia were just $\approx 0.3$ mag dimmer at maximum,
then a $\Lambda=0$ universe would be preferred (Riess et al.\ \cite{RFC98};
Leibundgut \cite{Lei00}).
Knowing whether the progenitors are SD or DD systems would constrain
the possible explosion models and improve the use of SNe~Ia as cosmological 
probes.

Several systematic radial velocity ({\bf RV}) searches for DDs have been
undertaken starting in the mid 1980's (e.g.\ Robinson \& Shafter \cite{RS87};
Bragaglia et al.\ \cite{BGR90}; Saffer, Livio \& Yungelson
\cite{SLY98}; 
Maxted \& Marsh \cite{MM99}).
Before 2001, combining all the surveys, $\approx 200$ white dwarfs 
have been checked for RV
variations with sufficient accuracy yielding
18 DDs with periods $P<6^{\rm d}.3$ (Marsh \cite{Mar00}; 
Maxted et al.\ \cite{MMM00}). 
None of the 18 systems seems massive enough to qualify as a SN~Ia
precursor. This is not surprising, as theoretical simulations suggest that
 only a few percent of all DDs are potential SN\,Ia progenitors
(Iben, Tutukov \& Yungelson \cite{ITY97}; Nelemans et al.\ \cite{NYP01}). 

In order to perform a definitive test of the DD scenario we have
embarked on a large spectroscopic survey  of 1500 white dwarfs using
the UVES spectrograph at the UT2 telescope (Kueyen) of ESO VLT to
search for RV variable white dwarfs (ESO {\bf S}N \,Ia
{\bf P}rogenitor surve{\bf Y} -- SPY). Details of our observations are
given in Sect.~\ref{s:observations}.
SPY will overcome the main limitation of all efforts so far to detect
DDs that are plausible SN~Ia precursors: the samples of surveyed
objects were too small.  

The SPY program has now just reached mid term, but already
dramatically increased the number of white dwarfs checked for
RV variations and the number of known DDs. Our sample
includes many short period binaries, several with masses closer to
the Chandrasekhar limit than any system known before. An overview of
our DD results is given in Sect.~\ref{s:DD}.

SPY produces a unique sample of high-quality, high-resolution spectra
of white dwarfs. This is a large, homogeneous database of enormous value for 
many other areas of research.  Spin-off opportunities in general and
projects started by us to exploit this treasure chest are  discussed 
in Sect.~\ref{s:spinoff}. 

\section{Observations and data reduction}
\label{s:observations}

\begin{figure*}
\resizebox{\hsize}{!}
{\includegraphics*[bb=20 245 590 750]{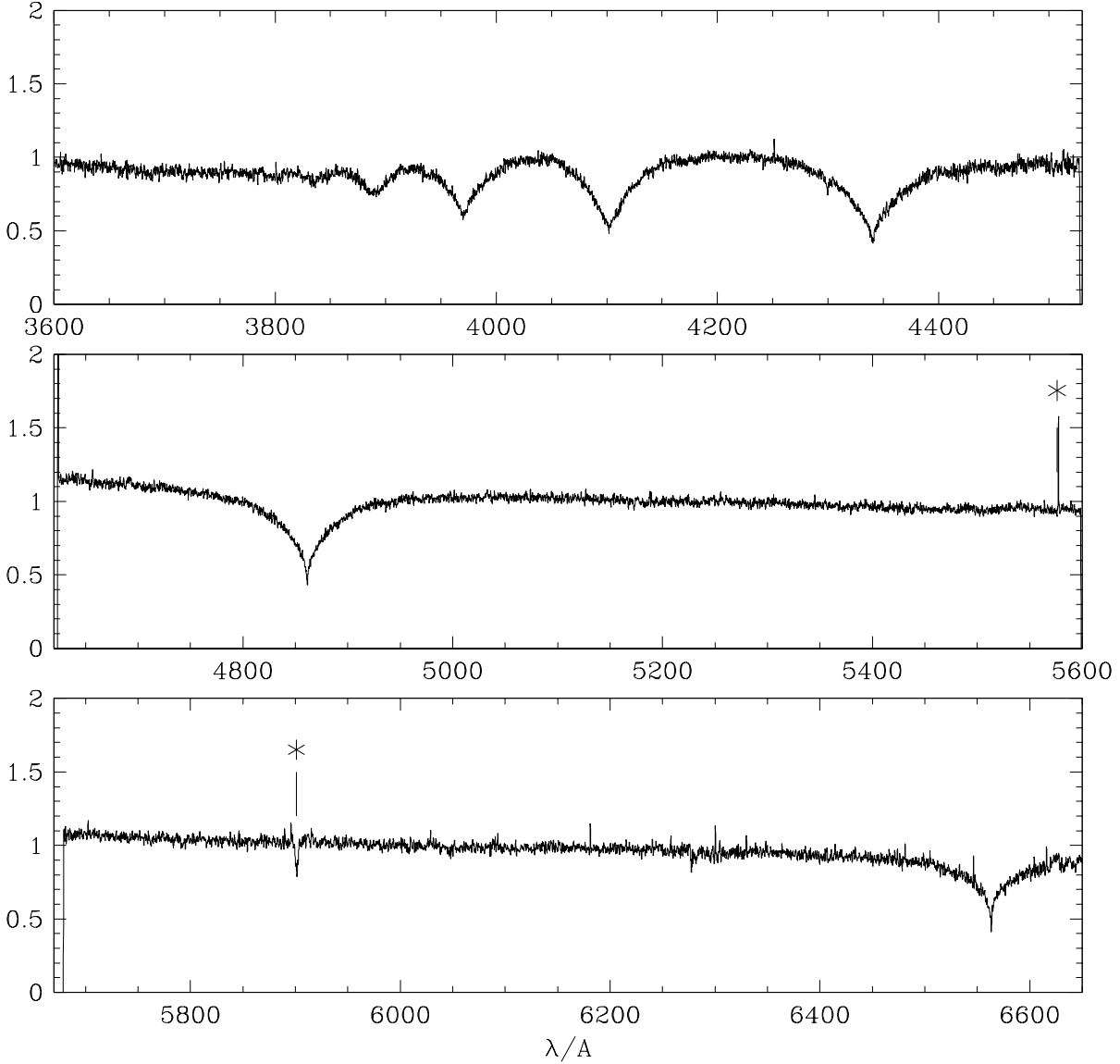}}
\caption{Pipeline reduced UVES spectrum of the DA white dwarf
HE\,0403$-$4129. The panels correspond to the blue channel and to both
parts of the red channel covered by different CCDs. Artifacts are
marked by asterisks. The spectrum was slightly smoothed and rebinned
as described in the text.}
\label{f:optical}
\end{figure*}

Targets for SPY are drawn from five sources: the white
dwarf catalog of McCook \& Sion (\cite{MS99}), the Hamburg ESO Survey
(HES; Wisotzki et al.\ \cite{WCB00}; Christlieb et al.\ \cite{CWR01}),
the Hamburg Quasar Survey (Hagen et al.\ \cite{HGE95}; Homeier et al.\
(\cite{HKH98}), the Montreal-Cambridge-Tololo survey (MCT; Lamontagne
et al.\ \cite{LDW00}), and the Edinburgh-Cape survey (EC; Kilkenny et
al.\ \cite{KOK97}). Our selection criteria were spectroscopic
confirmation as white dwarf (at least from objective prism spectra)
and $B\le 16.5$.

Spectra were taken with the UV-Visual Echelle Spectrograph (UVES) of
the UT2 telescope (Kueyen) of the ESO VLT. UVES
is a high resolution Echelle spectrograph, which can reach a
resolution of 110,000 in the red region with a narrow slit (cf.\
Dekker et al.\ \cite{DOK00} for a description of the instrument). Our
instrument setup (Dichroic~1, central wavelengths 3900\,\AA\ and
5640\,\AA ) uses UVES in a dichroic mode with a $2048\times 4096$ EEV
CCD windowed to $2048\times 3000$ in the blue arm, and two CCDs, a
$2048\times 4096$ EEV and a $2048\times 4096$ MIT-LL, in the red
arm. Nearly complete spectral coverage from 3200\,\AA\ to 6650\,\AA\
with only two $\approx$80\,\AA\ wide gaps at 4580\,\AA\ and 5640\,\AA\
is achieved.  In the standard setting used for our observations UVES
is operated with an $8''$ decker in the blue arm and an $11''$ decker
in the red arm.

Our program is implemented as a service mode program. It
takes advantage of those observing conditions, which are not usable by
most other programs (moon, bad seeing, clouds)
and keeps the VLT busy when
other programs are not feasible. A wide slit ($2.1''$) is used to
minimise slit losses and a $2\times 2$ binning is applied to the CCDs
to reduce read out noise. Our wide slit reduces the spectral
resolution to $R=18\,500$ (0.36\,\AA\ at H$\alpha$) or better, if
seeing disks were smaller than the slit width. Depending on
the brightness of the objects, exposure times of 5\,min or 10\,min
were chosen. The S/N per binned pixel (0.03\,\AA) of the extracted
spectrum is usually 15 or higher. 
Due to the nature of
the project, two spectra at different, ``random'' epochs separated 
by at least one day are observed.

Although our program is carried out during periods of less favourable
observing conditions, the seeing is often smaller than the selected
slit width of $2.1''$. This can, in principle, cause wavelength
shifts, if the star is not placed in the center of the slit. However,
since according to the standard observing procedure the star is first
centered on a narrow slit before the exposure with the broader slit is
started, it can be expected that the star is usually well centered in
the slit. We note that possible wavelength shifts can be corrected
with the telluric absorption features present in the red region of our
spectra. 

\paragraph{Data reduction:}
ESO provides a data reduction pipeline for UVES, based on MIDAS
procedures. The raw data are bias and interorder background subtracted
and extracted with an optimum extraction algorithm. The orders are
flatfielded, rebinned, and merged. Wavelength calibration is performed
by means of ThAr calibration lamp exposures. The quality of the
reduced spectra is in most cases very good; especially the removal of
the interorder sensitivity variation and merging of the orders works
very well. Sometimes the reduction pipeline produces artifacts of
varying strength, e.g.\ a quasiperiodic pattern in the red region
similar in appearance to a fringing pattern. In a few cases either the
blue or the red part of the spectrum has extremely strong artifacts of
unknown orign. A ``UVES'' context for MIDAS is available, which
provides reduction routines adapted for the UVES spectrograph and can
be used to re-reduce spectra. As an example of the quality achievable with the 
pipeline  reduction one spectrum without extraction problems is shown in 
Fig.~\ref{f:optical}.
 
Remaining large scale variations of the spectral response function
are removed by utilizing nearly featureless spectra of DC white
dwarfs or of sdO stars with narrow spectral lines, which are
normalised and used to derive an approximate response function. This
function was applied to the other spectra. Since the sampling of
the spectra is much higher than needed for many purposes (besides
measuring RVs), we produced a version of the spectra,
which were rebinned to 0.1\,\AA\
stepsize and smoothed with a Gaussian of 0.2\,\AA\ FWHM. This
produces only a slight degradation of the resolution while
considerably improving the signal-to-noise level.

\paragraph{Radial velocity measurement:}
Since SPY produces a large number of spectra, which have to be checked
for RV variations, a fast and reliable algorithm to measure RV shifts
is necessary. One often applied standard technique is cross correlation of
observed spectra. The RV shift is measured from the peak of the
correlation function. However, it turned out that this method
is not well suited for the determination of RVs 
from the narrow cores of the broad Balmer lines of DA white dwarfs. 
This can be understood as follows. If we limit the correlation to,
say, the inner 10\,\AA\ of the line to avoid spurious
signals caused by normalization problems etc., the boundaries of the
integration interval are located in the line wings. Contributions from both
interval boundaries (which generally do not cancel out) result
in a large
extra signal in addition to the real correlation signal, which makes
the determination of the correlation peak very difficult or even impossible. 

We overcome this problem by applying a correlation based on a 
$\chi^2$ test instead of the usual
cross correlation. Our procedure starts with a noise estimate from the
rms scatter  computed within a moving box. 
Cosmics etc.\ are flagged out with an clipping algorithm. Since it is important
for computing a realistic $\chi^2$ that the spectra have the same flux
level, we scale the second spectrum to the first one using a low order
polynomial to account for possible differences of the continuum slope. 
Afterwards $\chi^2$ is calculated as a function of RV
difference, by shifting the second spectrum in small RV steps
(0.1\,\kms)
and
calculate $\chi^2$ by direct numerical integration (without rebinning). Since
the calculation of $\chi^2$ for two noisy spectra results in some
noise in the resulting $\chi^2$ curve, finally a 
slight smoothing (Gaussian of FWHM 1\,\kms) is applied. 
The RV shift is evaluated from
the minimum $\chi^2$. Error margins can be estimated by determining
the intervals corresponding to a certain $\Delta \chi^2$ as described
in Press et al.\ (\cite{PTV92}).

From a formal point of view $\chi^2$ tests were invented to test
and find best fitting {\it models}. Nevertheless, 
our method works and produces   
reliable results with the observed spectra as well. 
We performed several tests: 
Two observed spectra were correlated with a model spectrum and the RV
shifts added, which yielded virtually the same result as the direct  $\chi^2$ 
correlation of both observed spectra. We also made 
experiments with simulated spectra, produced from model spectra by adding 
artificial noise and known RV shifts. These RV differences could
always be reproduced from the $\chi^2$ correlation within the
estimated error limits. 
One great advantage of our procedure is
its flexibility and that it can easily be applied to measure RV shifts
in stars of different spectral types (Balmer lines of DA white dwarfs,
He\,I lines of DBs, and metal lines of sdB stars like HE\,1047$-$0436
discussed below).
We routinely measure RVs with an accuracy of $\approx 5$\,\kms\
or better, therefore running only a very small risk of missing a merger
precursor, which have orbital velocities of 150\,km s$^{-1}$ or higher.

\section{Results on double degenerates}
\label{s:DD}

\begin{figure}
\resizebox{\hsize}{!}
{\includegraphics*[bb=20 40 590 740]{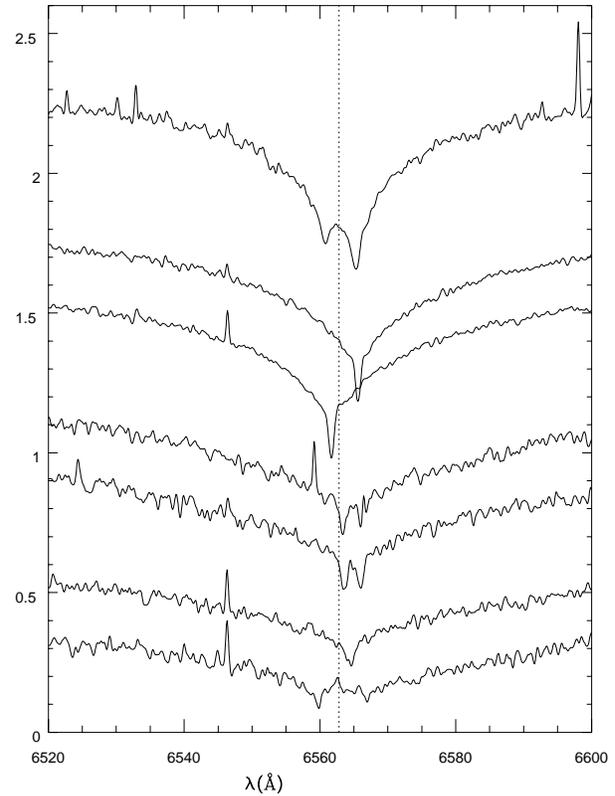}}
\caption{Three double-lined and one single lined RV variable DDs from
our VLT survey. The dotted line marks the rest wavelength of H$\alpha$.
The spectra are rebinned and smoothed to a resolution of $\approx$0.5\,\AA.}
\label{f:halpha}
\end{figure}

We
have analysed spectra of 387 white dwarfs taken during the first 15 months of
the SPY project and detected 54 new DDs, 7 are double-lined
systems (only 6 were known before). The great advantage of
double-lined binaries is that they provide us with a well determined
total mass. Since it is likely that the SPY sample contains even more
double-lined systems (with a faint secondary), 
we will check follow-up observations of apparently
single-lined systems for the signature of the secondary.
Our sample includes many
short period binaries (some examples are discussed in
Sect.~\ref{s:followup}), several with masses closer to the
Chandrasekhar limit than any system known before. 
In addition, we detected 8 RV
variable systems with a cool main sequence companion (pre-cataclysmic
variables; pre-CVs). Some examples of single-lined and double-lined
DDs are shown in Fig.~\ref{f:halpha}. A more detailed inventory of the
SPY sample is given in
Table~\ref{t:DDs}.
Our observations have already increased the DD sample
by a factor of four.  After completion, a final sample of 150 to 200 DDs
is expected.

\begin{table}
\caption{Fraction of RV variable stars in the current SPY sample for 
different spectral classes. WD+dM denotes systems for which a
previously unknown cool companion is evident from the red spectra.}
\label{t:DDs}

\begin{tabular}{l|r|r|r}
Spectral type	&total	&RV variable	&detection rate\\ \hline
All	&387	&62	&16\%\\
DA	&310	&53	&17\%	\\
non-DA (DB,DO,DZ)	&53	&1	&2\%	\\
WD+dM	&24	&8	&33\%	\\
\end{tabular}
\end{table}

Follow-up observations of this sample are mandatory to exploit its
full potential. Periods and white dwarf parameters
must be determined to find potential SN\,Ia progenitors among the candidates. 
Good statistics of a large DD sample will also set stringent
constraints on the evolution of close binaries, which will
dramatically improve our understanding of this phase of stellar
evolution.  
During our follow-up observations we have detected a very promising
potential SN\,Ia precursor candidate. However, some additional spectra
are necessary to verify our RV curve solution. Results will be
reported elsewhere.

Although important information like the periods, which can only
be derived from follow-up observations, are presently lacking for most
of the stars, the large sample size already allows us to draw some
conclusions. (Note that fundamental white dwarf parameters like masses are
known from the spectral analysis described below). One interesting
aspect concerns white dwarfs of non-DA classes (basically the helium-rich
spectral types DB, DO, and DZ, 
in contrast to the hydrogen-rich DAs).  SPY is the
first RV survey which performs a systematic investigation of both
classes of white dwarfs: DAs {\it and} non-DAs. Previous surveys were restricted
to DA
white dwarfs, because the sharp NLTE core of H$\alpha$ provides a very
accurate RV determination. This feature is not present in non-DA (DB,
DO) spectra, but the use of several helium lines enables us to reach a
similar accuracy.  Thus the low number of DDs found for the non-DA
spectral types is statistically significant (Table~\ref{t:DDs}) and
indicates that important evolutionary channels for the formation of
DDs produce preferentially hydrogen-rich white dwarfs of the DA variety.

\subsection{Highlights of our follow-up results}
\label{s:followup}

Follow-up observations are necessary to determine the system
parameters of the DDs. We concentrated on candidates with high
RV variations, indicating short periods, because the probability to
find
potential SN\,Ia candidates among these systems is highest. 
However, let us note that probably
some of the ``small $\Delta$RV'' DDs could be short period systems 
(possibly even SN\,Ia progenitors) with low inclination angles and/or
unfavourable phase differences of the SPY observations. 
Follow-up observations have been carried out at the VLT as well as 
with the 3.5\,m telescope of the Calar
Alto observatory/Spain. 

In the following paragraphs we present results for two of our most
interesting binaries: the subdwarf\,B + white dwarf system HE\,1047--0436
(Napiwotzki et al.\ \cite{NEH01}) 
and the double-lined DA+DA binary HE\,1414--0848 (Napiwotzki et al.\
\cite{NKN01})  with a mass only 10\%
below the Chandrasekhar limit. 

\begin{figure}
\resizebox{\hsize}{!}
{\includegraphics*[bb=29 40 592 740]{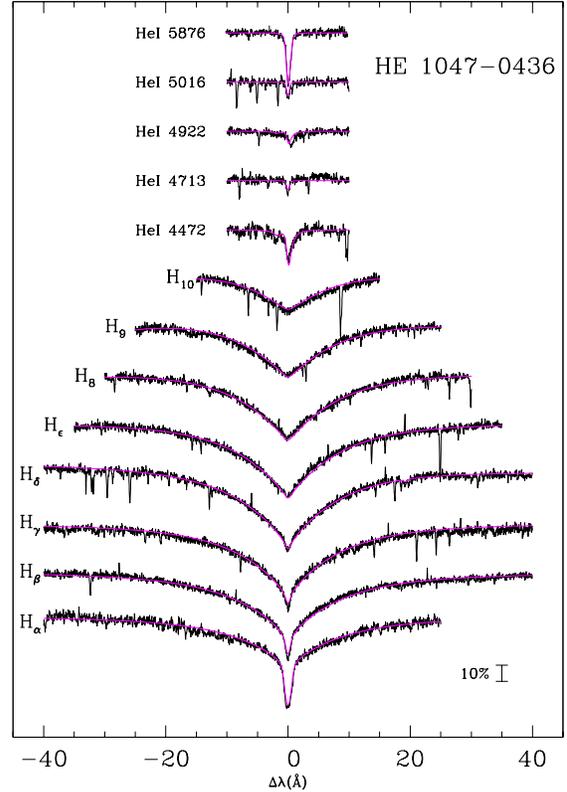}}
\caption{Fit of the hydrogen and helium lines of the RV corrected, coadded UVES
spectra of HE\,1047--0436.}
\label{f:HE1047fit}
\end{figure}

\paragraph{HE\,1047--0436:}
HE\,1047--0436 was discovered by the Hamburg ESO Survey (HES) as a
potential hot white dwarf and, therefore, included in our survey. The UVES
spectra (Fig.~\ref{f:HE1047fit})
showed that it is in fact a subluminous B star (sdB) 
with a rather large RV shift of 160\,\kms, which made the star
a prime target for further study. SdB stars are pre-white dwarfs of
low mass ($\approx 0.5 M_\odot$) still burning helium in the core,
which will evolve directly to the white dwarf stage omitting a second red giant
phase. 

Recently, sdBs with white dwarf components have been proposed as potential
SNe\,Ia progenitors by Maxted et al.\ (\cite{MMN00}) who discovered that
KPD\,1930+2752 is a sdB+WD system. Its total mass exceeds the
Chandrasekhar mass and the components will merge within a Hubble time,
which makes KPD\,1930+2752 a SN\,Ia progenitor candidate (although
this interpretation has been questioned recently, Ergma et al.\ \cite{EFY01}).

\begin{figure}
\resizebox{\hsize}{!}
{\includegraphics*[bb=0 00 740 550]{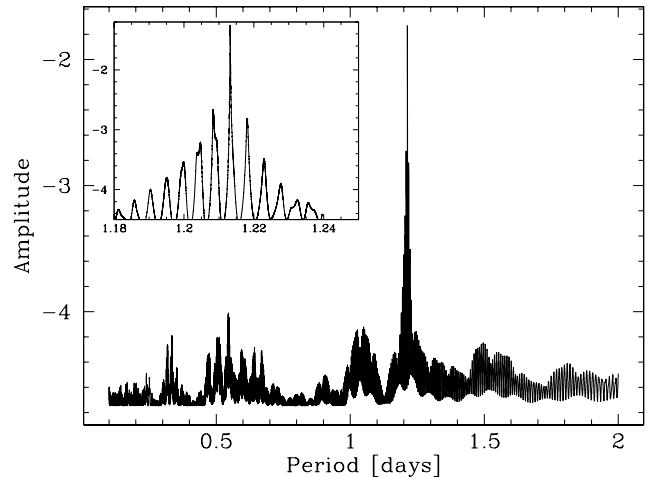}}
\caption{Power spectrum of the HE\,1047--0436 measurements used for
the periodogram analysis.}
\label{f:HE1047pow}
\end{figure}

\begin{figure}
\resizebox{\hsize}{!}
{\includegraphics*[bb=0 0 740 550]{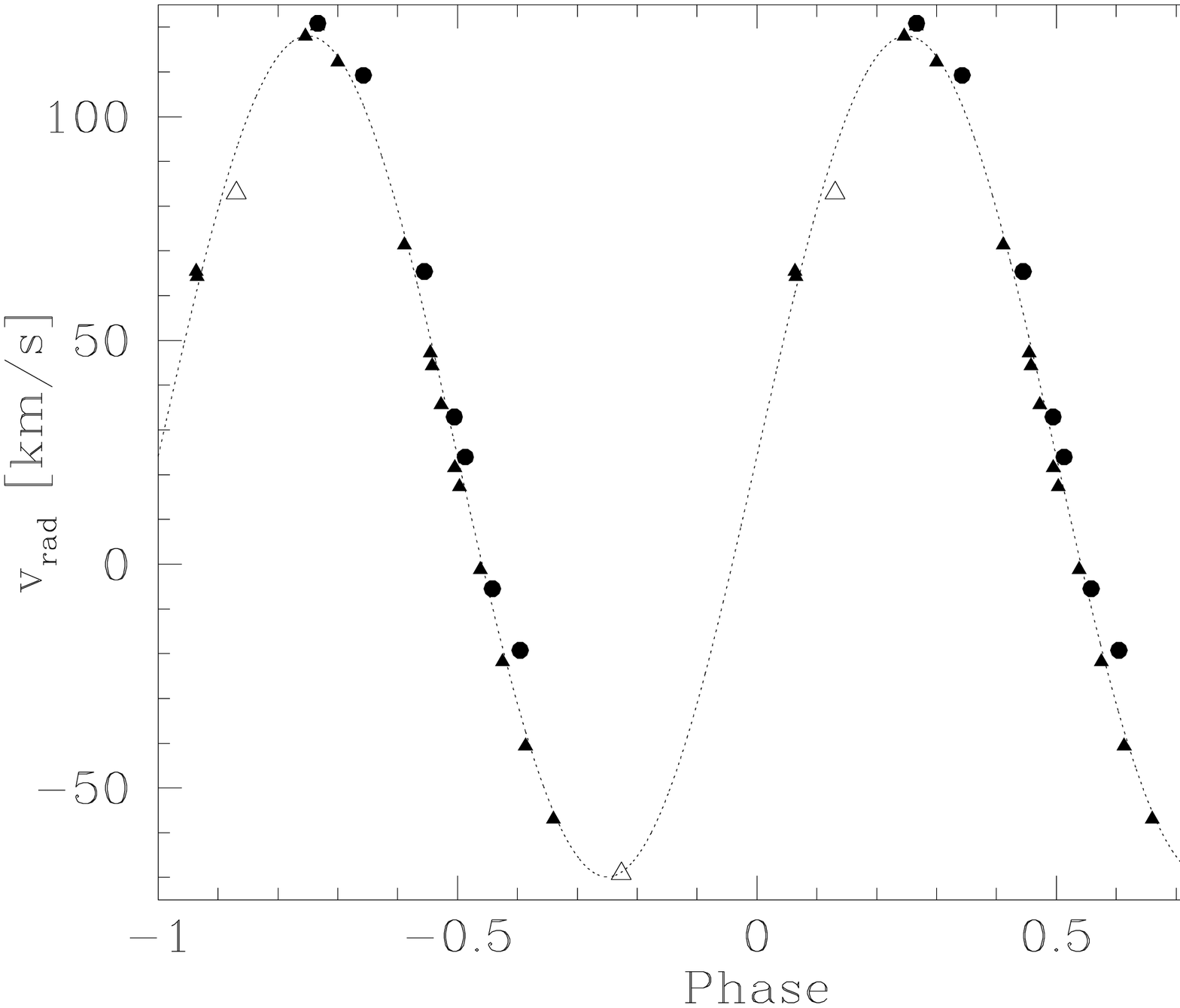}}
\caption{Measured RVs as a function of orbital phase and fitted sine
curve for HE\,1047--0436.}
\label{f:HE1047RV}
\end{figure}

The high RV shift of HE\,1047--0436 prompted us to perform 
follow-up observations (Napiwotzki et al.\ \cite{NEH01}). The orbital 
period of $P = 29^{\mathrm{h}} 7^{\mathrm{m}} 5^{\mathrm{s}}$,
 a semi-amplitude of 94\,km\,s$^{-1}$, and a
minimum mass of the invisible companion of $0.44\,M_\odot$ are derived
from the analysis of the RV curve (Figs.~\ref{f:HE1047pow}
and~\ref{f:HE1047RV}).
We use an upper limit
on the projected rotational velocity of the sdB star to constrain the
system inclination and the companion mass to $M>0.71\,M_\odot$,
bringing the total mass of the system closer to the Chandrasekhar
limit.  However, the system will merge due to loss of angular momentum
via gravitational wave radiation only after several Hubble times.
Atmospheric parameters ($T_{\mathrm{eff}} = 30200$\,K, $\log g = 5.7$) and 
metal abundances are also derived (cf.\ Fig.~\ref{f:HE1047fit}).
The resulting values are typical for sdB stars (cf.\ e.g.\ Edelmann et al.\
\cite{EHN02}).

\paragraph{HE\,1414--0848:}

\begin{figure}
\resizebox{\hsize}{!}
{\includegraphics*[bb=20 40 590 750]{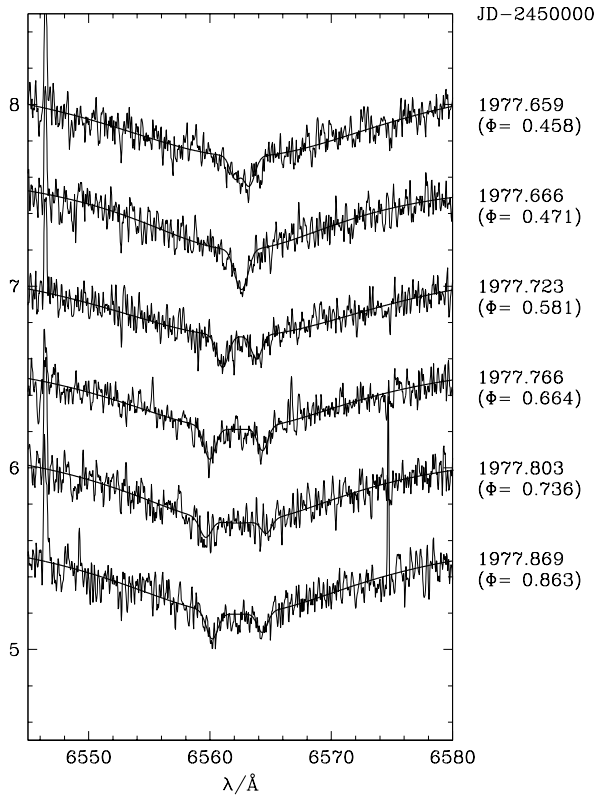}}
\caption{H$\alpha$ spectra of HE\,1414-0848 
covering 5 hours during one night
together with a fit of the line cores. The numbers indicate the Julian
date of the exposures and the orbital phase $\phi$.}
\label{f:HE1414ha}
\end{figure}

\begin{figure}
\resizebox{\hsize}{!}
{\includegraphics*[bb=00 20 740 550]{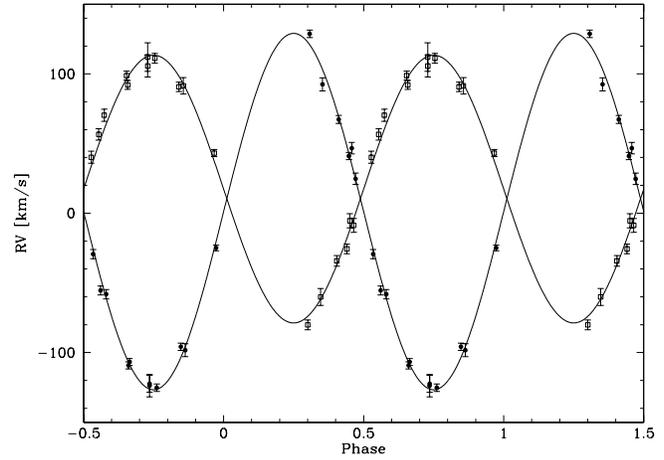}}
\caption{Measured RVs as a function of orbital phase and
fitted sine curves for HE\,1414-0848. Filled circles/open rectangles 
 indicate the less/more massive component. Note the difference of the
``systemic velocities'' $\gamma_0$ between both components caused by
gravitational redshift.}
\label{f:HE1414rv}
\end{figure}

HE\,1414$-$0848 is a double-lined DA+DA binary (Fig.~\ref{f:HE1414ha}).  
The orbital period of
$P = 12^{\mathrm{h}} 25^{\mathrm{m}} 44^{\mathrm{s}}$ and
semi-amplitudes of 127\,km\,s$^{-1}$ and 96\,km\,s$^{-1}$ are derived
for the individual components (Napiwotzki et al.\ \cite{NKN01}).  
RV curves for both
components are displayed
in Fig.~\ref{f:HE1414rv}. 
The ratio of velocity amplitudes is directly related to the
mass ratio of both components. Additional information comes from the 
mass dependent gravitational redshift 
$z =\Delta \lambda/\lambda= GM R^{-1}c^{-2}$, which for 
 a given mass-radius relation
can be computed as a function of white dwarf mass. 
The difference in gravitational redshift
corresponds to the apparent difference 
of ``systemic velocities'' of both components, as derived from the RV
curves.
Only one set of individual 
white dwarf masses fulfils the constraints given by both the amplitude
ratio and redshift difference.
We estimate the masses of
the individual components with this method: 
$0.54M_\odot$ and $0.72M_\odot$.  Another
estimate of the white dwarf parameters is available from a model
atmosphere analysis of the combined spectrum. The formal result
indicates an average mass of $0.68M_\odot$ consistent with the
results derived from the analysis of the RV curve.  The
total mass of the HE\,1414$-$0848 system is $1.26M_\odot$, only 10\%
below the Chandrasekhar limit!  The system will merge due to loss of
angular momentum via gravitational wave radiation after two Hubble
times.  Therefore HE\,1414$-$0848 does not qualify as a SN\,Ia
progenitor, but it is the most massive short period DD 
known today.

\section{Spin-off results}
\label{s:spinoff}

SPY produces an immense, unique
sample of very high resolution white dwarf spectra. This database will
have a large impact on many fields of white dwarf science. It will allow us for
the first time to tackle many longstanding questions on a firm
statistical basis.  Among those are the mass distribution of white dwarfs,
kinematical properties of the white dwarf population, surface compositions,
luminosity function, rotational velocities, and detection of weak
magnetic fields.
We have therefore decided to make these results available to the
community in the form of a catalog with some preliminary
interpretation as soon as feasible, not waiting for the completion of
the SPY project. A first part of this catalog was published in the
recent paper of Koester et al.\ (\cite{KNC01}), covering observations of about
200 white dwarfs of spectral types DA and DB. 
For all spin-off opportunities mentioned above the
statistics will be dramatically improved by the final white dwarf spectra
database. 
We are 
exploiting the
SPY sample for two spin-off projects, which take advantage of the high
spectral resolution: the kinematics of white dwarfs and 
their rotational velocities.

\paragraph{Basic parameters and mass distribution of white dwarfs:}
The response of the UVES spectrograph is sufficiently stable and well
behaved to allow the determination of temperatures, gravities, and
masses for the programme white dwarfs.  Thus we carried out a model atmosphere
analysis and published the results in our catalog paper (Koester et
al.\ \cite{KNC01}). Agreement of our results with literature values is quite
good for the DA sample, thus confirming that the UVES spectra can be
used for accurate and reliable parameter determination of white dwarfs. This
was not necessarily expected, since the individual Balmer lines in the
DA span several echelle orders.  Note that our catalog paper included
the second ever published analysis of a sample of helium-rich DB white dwarfs.

Our parameter determinations 
allow the selection of special types of white dwarfs for
further follow-up investigations (by us as well as by other members of
the community), and are a prerequisite for the determination of space
velocities (see next paragraph).

\paragraph{Kinematical properties of white dwarfs:}
The large SPY sample of white dwarfs can be used  to study the
kinematics of these stars in our Galaxy, especially with respect to
their belonging to the thin or thick disk population.  The aim is to
verify recent claims that a substantial fraction of the mass of the
Galactic disk may be provided by thick disk white dwarfs, as it is concluded
from the abundance analysis of a local sample of F and G type main
sequence stars (Fuhrmann \cite{Fuh01}).


\begin{figure}
\resizebox{\hsize}{!}
{\includegraphics*{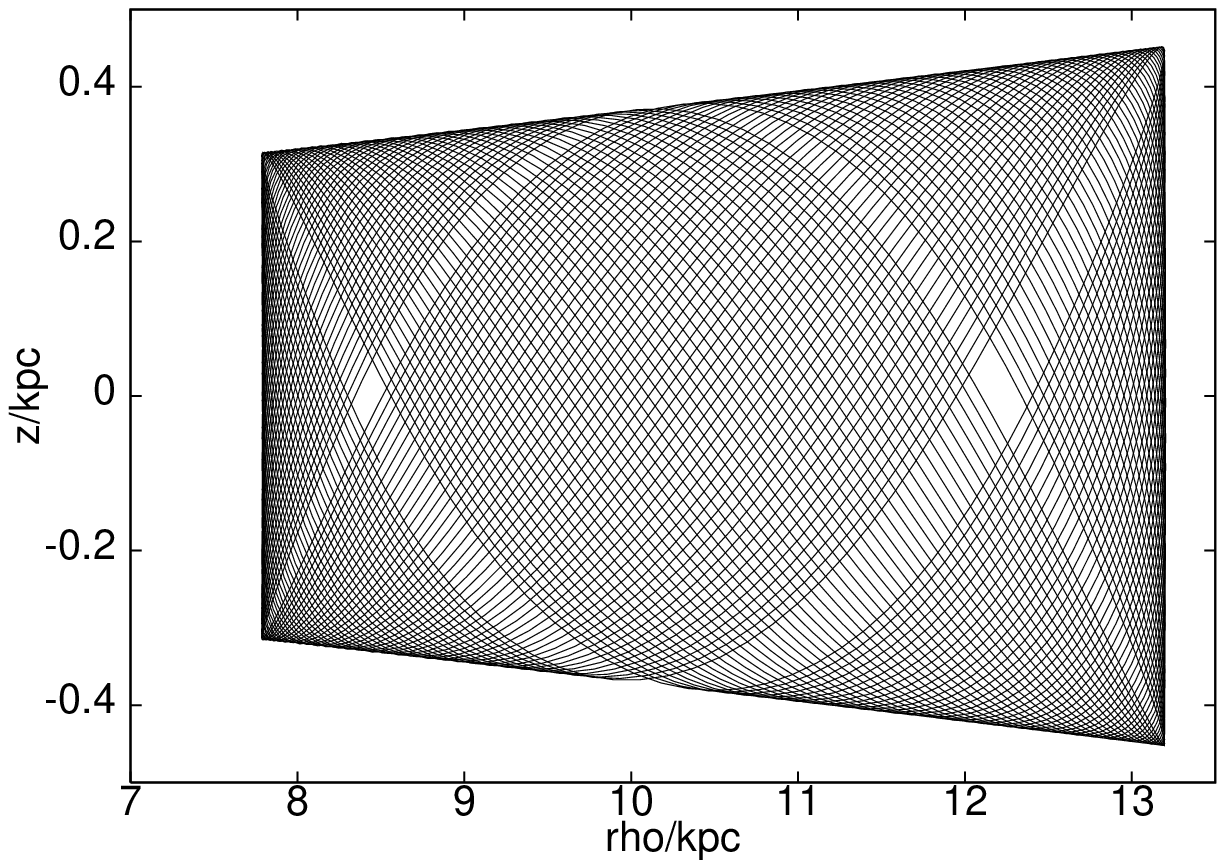}}
\resizebox{\hsize}{!}
{\includegraphics*{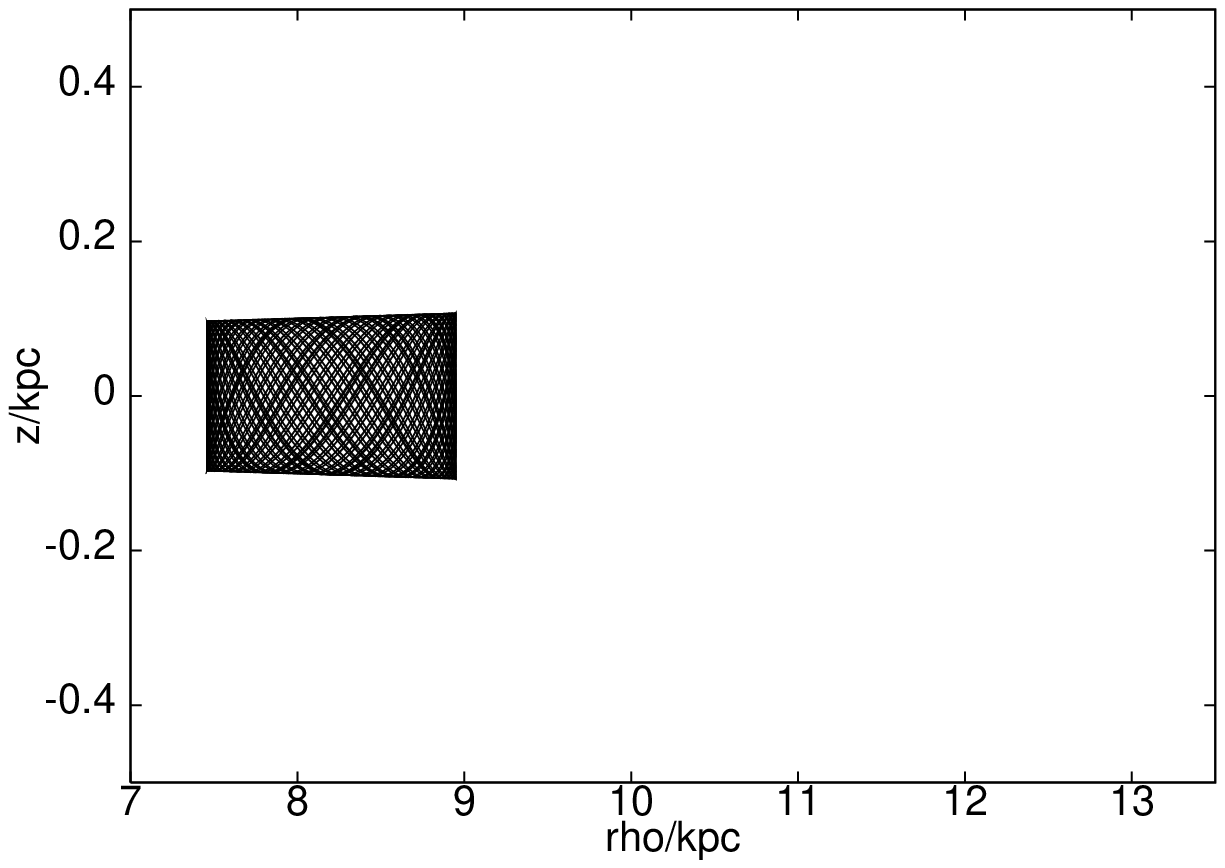}}
\caption{Meridional projection of the Galactic orbits of WD\,0013$-$241 (top)
and WD\,0016$-$258 (bottom) computed over 10\,Gyr. WD\,0016$-$258 is a 
typical thin disk star with an orbit of small eccentricity reaching only small
heights above the Galactic plane. The orbit of WD\,0013$-$241 on the
other hand has high eccentricity and reaches larger distances from the
Galactic
plane,
typical for a thick disk star.}
\label{f:orbit}
\end{figure}

We combined the measured RVs of the white dwarfs 
and determined proper motions from the DSS1/DSS2 and USNO
data. Distances and corrections of the RV for gravitational 
redshift are computed from the fundamental parameters 
derived by Koester et al.\ (\cite{KNC01}). 
Thus the space motion
in three dimensions is known and Galactic orbits can be computed 
from a code developed by Odenkirchen \& Brosche (\cite{OB92}). 
Examples of thin and thick
disk orbits are shown in Fig.~\ref{f:orbit}.
So far 47 white dwarfs have been analysed.  This will be the first
kinematic study of a large sample of white dwarfs, which is based on
the complete set of space motion in three dimensions.

\paragraph{Rotational velocities:}

High resolution spectra of the H$\alpha$ line are very useful for the
determination of rotational velocities ($v \sin i$)
 of white dwarfs, which puts
constraints on the evolution of angular momentum during stellar
evolution. Two recent studies (Heber et al.\ \cite{HNR97}, 
Koester et al.\ \cite{KDW98}) 
increased the number of determinations just to 54. 
It turned out that white dwarfs are generally very slow rotators with
a few, puzzling exceptions.

\begin{figure}
\resizebox{\hsize}{!}
{\includegraphics*[bb=10 10 740 550]{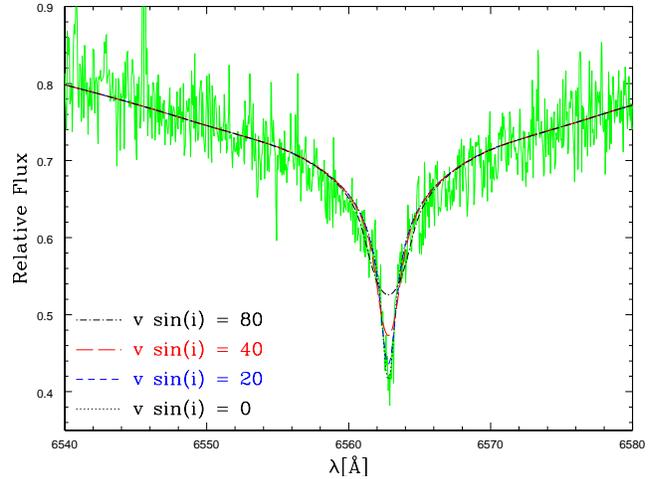}}
\caption{Observed core of the H$\alpha$ line compared to synthetic
spectra computed for different rotational velocities $v \sin i$. This
spectrum allows to put  an upper limit of {15\,\kms} on $v \sin i$.}
\label{f:vrot}
\end{figure}

Many spectra
taken for the SPY project allow the determination of rotational
velocities with {\it better} accuracy than previous studies. We
started the systematic analysis of the present sample (one example
is displayed in Fig.~\ref{f:vrot}). The final
sample will dramatically increase the number of white dwarfs with known
rotational velocities and allow for the first time to explore trends
of $v \sin i$ with, e.g., cooling age.

\paragraph{Chemical abundances:}
High-resolution spectra are very valuable for the determination of surface 
compositions. Most white dwarfs have very pure atmospheres containing only 
hydrogen (DA) or helium (DO, DB). However, in the atmospheres of some hotter 
white dwarfs as well as in a fraction of cooler white dwarfs (types 
DZ, DAZ, DBZ) 
metal lines can be detected. Our understanding of this phenomenon is currently 
limited by the insufficient sample size of white dwarfs observed with high
enough resolution. The largest published sample contains 38 white dwarfs 
observed with the HIRES spectrograph of the Keck\,I telescope 
(Zuckerman \& Reid \cite{ZR98}). 
It is obvious that SPY can help to improve the 
statistics by a large factor.

\begin{figure}
\resizebox{\hsize}{!}
{\includegraphics*[angle=-90, bb=50 10 570 740]{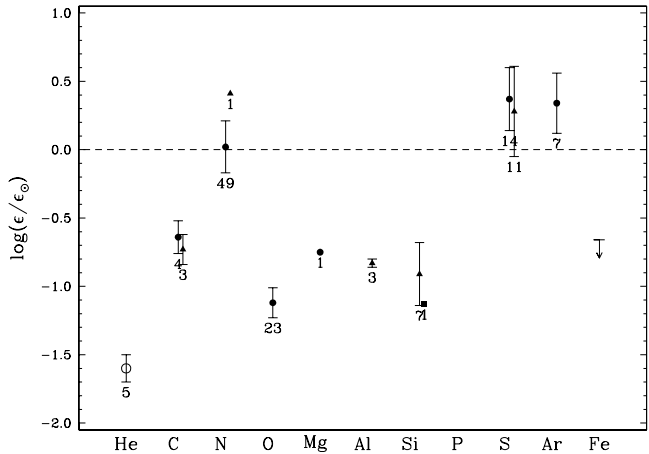}}
\caption{Chemical abundances in the atmosphere of the subdwarf\,B star 
HE\,1047$-$0436 compared to the solar values. Numbers indicate the number
of lines measured for each ionisation stage.}
\label{f:he1047metall}
\end{figure}

A few of the white dwarf candidates observed by SPY turned out to be 
subdwarf~B stars (i.e.\ pre-white dwarfs with lower surface gravities). 
These stars display puzzling abundance patterns likely caused by diffusion
processes (cf.\ Edelmann et al.\ \cite{EHN02}). 
As an example for the resulting chemical abundance pattern
we present in 
Fig.~\ref{f:he1047metall} our analysis of the binary
HE\,1047$-$0436 discussed above (Sect.~\ref{s:followup}).

\section{Summary}

SPY has now reached about mid term and it has already 
tripled the number of white dwarfs
checked for RV variablity (from 200 to 600) and
quadrupled the number of known DDs (from 18 to 72) compared to the
results achieved during the last 20 years. 
Our sample includes many
short period binaries (Napiwotzki et al.\ \cite{NEH01}, \cite{NKN01}), 
several with masses closer to the
Chandrasekhar limit than any system known before, greatly improving the
statistics of DDs. 
We expect this survey to produce a sample of 150 to 200
DDs. 

This will allow us not only to find several of the long sought potential
SN~Ia
precursors (if they are DDs), but will also provide a census of the
final binary configurations, hence an important test for the theory of
close binary star evolution after mass and angular momentum losses
through winds and common envelope phases, which are very difficult to model.
An empirical calibration provides the 
most promising approach. A large sample of binary white dwarfs 
covering a wide range
in parameter space is the most important ingredient for this task.

SPY produces a unique sample of white dwarf spectra with many spin-off
opportunities, which will have a large
impact on white dwarf science.
We published a model atmosphere analysis of a first set of 200 white
dwarfs (Koester et al.\ \cite{KNC01}). This allowed us to construct a mass 
distribution of a large sample of white dwarfs and made this sample
known to the community for further follow-up investigations. We are 
exploiting the
SPY sample for two spin-off projects, which take advantage of the high
spectral resolution: we study the kinematics of white dwarfs and
determine their rotational velocities.
Members of the community interested in spin-off opportunities are invited
to participate in the exploitation of the SPY sample.

\acknowledgements 
We express our gratitude to the ESO staff, for providing invaluable
help and conducting the service observations and pipeline reductions,
which have made this work possible. We gratefully acknowledge the assistence
of the Calar Alto staff.
D.K.\ and D.H.\ thank the Deutsche
Forschungsgemeinschaft (DFG) for their support (Ko\,739/10-3). C.K.\
and E.-M.P. also acknowledge support by the DFG
(Na\,365/2-1). S.M. was supported by a grant (50\,OR\,96029-ZA) from
the Bundesministerium f\"ur Bildung und Forschung through the DLR.
L.Y.\ is supported by RFBR grant
99-02-16037 and ``Program Astronomy'' grant 1.4.4.1. This work was
supported by DFG travel grants Na\,365/3-1 and Na\,365/4-1.

\end{document}